\documentclass{PoS}
\bibliography{/home/dave/conferences/como/}
\def\o3{[O~\small III\normalsize ]}
\def\s2{[S~\small II\normalsize ]}

\title{Jet-Powered Optical Nebulae From X-ray Binaries}

\ShortTitle{Jet-Powered Optical Nebulae From X-ray Binaries}

\author{\speaker{David M. Russell}%
	 \\
        University of Southampton, UK\\
        E-mail: \email{davidr@phys.soton.ac.uk}}

\author{Rob P. Fender\\
 University of Southampton, UK\\
 E-mail: \email{rpf@phys.soton.ac.uk}}

\author{Elena Gallo\\
 University of California, USA\\
 Chandra Fellow\\
 E-mail: \email{elena@physics.ucsb.edu}}

\author{James C. A. Miller-Jones\\
 University of Amsterdam, NL\\
 E-mail: \email{jmiller@science.uva.nl}}

\author{Christian R. Kaiser\\
 University of Southampton, UK\\
 E-mail: \email{crk@soton.ac.uk}}

\abstract{Accreting black holes and neutron stars release an unknown fraction of the infalling particles and energy in the form of collimated jets. The jets themselves are radiatively inefficient, but their power can be constrained by observing their interaction with the surrounding environment. Here we present observations of X-ray binary jet-ISM interactions which produce optical line emission, using the ESO/MPI 2.2~m and Isaac Newton Telescopes. We constrain the time-averaged power of the Cyg X--1 jet-powered nebula, and present a number of new candidate nebulae discovered. Comparisons are made to the large scale lobes of extragalactic AGN. We also speculate that some emission line emitters close to X-ray binaries in M31 are likely to be microquasar jet-powered nebulae.}

\FullConference{VI Microquasar Workshop: Microquasars and Beyond\\
September 18-22, 2006\\
Como, Italy}

\begin{document}

\section{Introduction}

Microquasars release an unknown amount of energy and matter into the interstellar medium (ISM) in the form of collimated jets (see \cite{fend06} for a review). Measuring as accurately as possible the jet power is key to understanding the physics of the accretion process and the matter and energy input from microquasars into the ISM. The most common method to estimate the power in the jets is from the core radio emission; a method requiring assumptions about the jet spectrum and radiative efficiency. The radiative efficiency is low and uncertain: typically estimated to be $\sim 5$ percent \cite{ogleet00,fend01}. The radiative power of the jets is dominated by the higher energy photons; the high-frequency break of the flat (spectral index $\alpha \sim 0$ where $F_{\nu}\propto \nu^{\alpha}$) optically thick part of the jet spectrum is likely to be at $\sim 2 \mu$m, at least for black hole X-ray binaries at high luminosities ($L_{\rm X}\ge 10^{36}$ erg s$^{-1}$) in the hard state \cite{corbfe02,russet06a} but is poorly constrained at lower luminosities.

Alternatively, the power of microquasar jets can be constrained from observing their interaction  with the surrounding medium. For this latter method, assumptions about the jet spectrum and radiative efficiency are not required, and the technique is widely practiced for FR II radio galaxies powered by active galactic nuclei (AGN).

\section{The Cyg X--1 jet-powered nebula}

The work in this section is described in more detail in a forthcoming paper \cite{russet06b}. Recently, a radio shell has been found in alignment with the resolved radio jet of the high-mass X-ray binary and microquasar Cygnus X--1 \cite{gallet05}. The shell was also observed to be bright in optical H$\alpha$ line emission. As a result, the radio emission was interpreted as originating from bremsstrahlung from the shock-compressed gas just behind the shock front, as is expected in models of jet-ISM interactions \cite{kaiset04}. Bremsstrahlung X-ray radiation has been seen from intracluster gas that is shock-compressed by the jets of AGN \cite{wilset06}. From the radio luminosity and constrained shock temperature and velocity, the time-averaged power in the Cyg X--1 jet was estimated by modelling the shell as a radiative shock wave: $9\times 10^{35}\leq P_{\rm Jet}\leq 10^{37}$ erg s$^{-1}$. Radio lobes from the synchrotron plasma within the bubbles have been seen associated with the jets of a few microquasars, for example 1E 1740.7--2942 \cite{miraet92} and GRS 1758--258 \cite{martet02}, but this is the first time the thermal shell surrounding these radio lobes has been identified for a microquasar.

\begin{figure}
\includegraphics[width=15cm,angle=0]{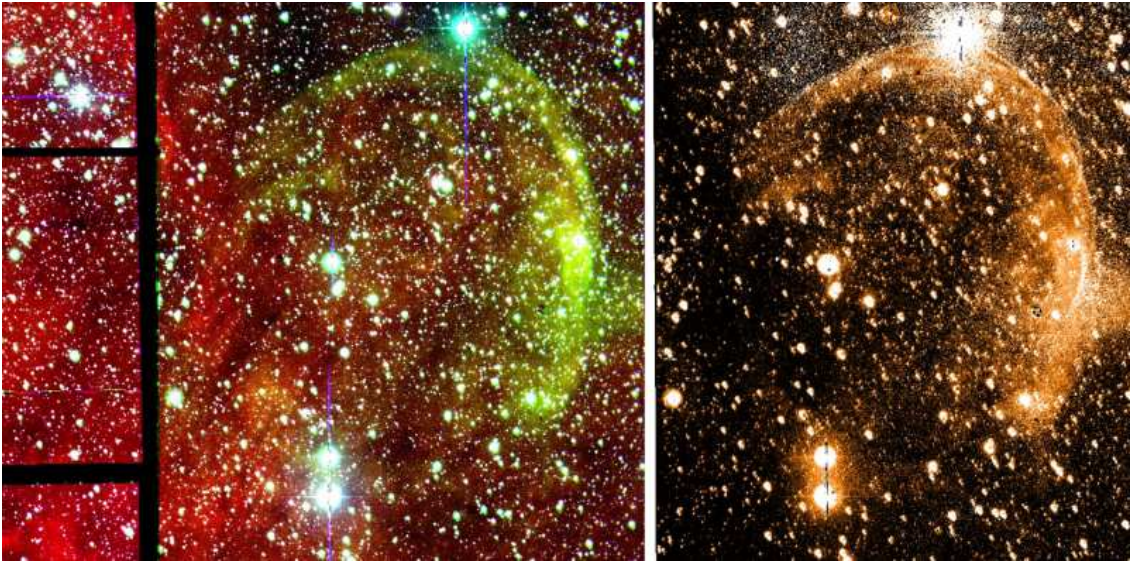}
\caption{Aligned and stacked emission line images of the Cyg X--1 nebula. North is up and east is to the left. Cyg X--1 itself is the brightest star near the bottom of each image. Left panel: Colour image (V-band continuum is blue, H$\alpha$ is red and [O III] 5007\AA~is green) illustrating the difference in [O III] surface brightness between the nebula and the nearby H II region to the east. Right panel: [O III] / H$\alpha$ image, revealing a thin luminous outer shell to the nebula at the shock front.}
\end{figure}

\subsection{Constraining the jet power from new optical emission line imaging observations}

We performed deep optical narrowband observations of Cyg X--1 and its nebula with the Wide Field Camera (WFC) on the Isaac Newton Telescope (INT). The data were reduced, stacked and mosaiced using the package \small THELI \normalsize \cite{erbeet05} and flux calibrated using \small IRAF\normalsize. The left panel of Fig. 1 is a colour image of the region comprising V-band continuum (blue), H$\alpha$ (red) and \o3 (5007\AA; green) line emission. The flux calibrated \o3 / H$\alpha$ ratio is presented in the right panel of Fig. 1. The bright thin outer shell in this image is not expected for photoionized gas, which should be diffuse. Shock-excited gas on the other hand is expected to have a shock front where the gas is ionized and compressed, and line ratios such as \o3 / H$\alpha$ are sensitive to the temperature and velocity of the shocked gas \cite{coxra85}. Non-radiative shock waves produce Balmer optical emission lines only \cite{raymet83}, so the existence of the strong \o3 line confirms the radiative nature of the Cyg X--1 nebula shock. The observed shell of enhanced \o3 / H$\alpha$ flux ratio is the first detection of the bow shock front from a shock wave powered by an X-ray binary jet.

\begin{figure}
\begin{center}
\includegraphics[width=10cm,angle=0]{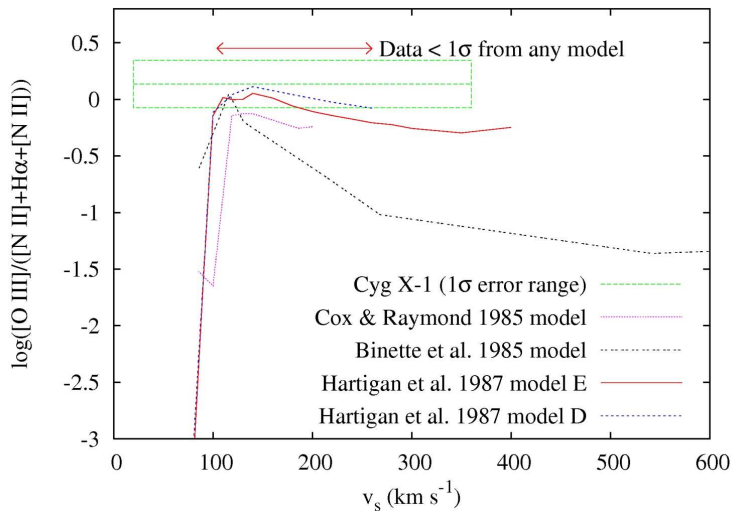}
\caption{The [O III] / H$\alpha$ ratio as a function of shock velocity predicted from four radiative shock models. The dashed rectangle represents the constrained range of values measured for the Cyg X--1 shock front \cite{gallet05}. The INT [O III] filter includes both lines of the doublet; 4958.9\AA~and 5006.9\AA~and the H$\alpha$ filter includes the two [N~II] lines at 6548.1\AA~and 6583.4\AA~in addition to 6562.8\AA~H$\alpha$, within its bandpass. We therefore plot log($F_{\rm [O~III]+[O~III]}$/$F_{\rm [N~II]+H\alpha +[N~II]}$) versus velocity.}
\end{center}
\end{figure}

From the flux calibrated \o3 / H$\alpha$ image (Fig. 1 right panel) we measured $F_{\rm \o3}/F_{\rm H\alpha}\sim 1.4$ for the bow shock front. According to models for radiative shocks, the preshock gas becomes ionized in a small shock velocity range: $90\leq v_{\rm s}\leq 120$ km s$^{-1}$ \cite{shulmc79}. \o3 is a high ionization line and is not significant at $v_{\rm s}\leq 90$ km s$^{-1}$. Using a number of advanced radiative shock models \cite{coxra85,bineet85,hartet87}, the observed \o3 / H$\alpha$ flux ratio can only originate in a shock wave with velocity $100\leq v_{\rm s}\leq 260$ km s$^{-1}$ ($1\sigma$) (Fig. 2). Using the methodology of Gallo et al. (2005) \cite{gallet05}, this corresponds to a gas temperature of $2.6\times 10^5 \leq T_{\rm s}\leq 1.6\times 10^6$ K and a refined time-averaged jet power to produce the nebula, of $4.1\times 10^{36}\leq P_{\rm Jet}\leq 1.0\times 10^{37}$ erg s$^{-1}$.

\subsection{Searching for ISM interactions with the counter jet}

There is no obvious nebula caused by the southern jet of Cyg X--1. The density of the ISM may be much less to the south, and the jet should travel undetected until being decelerated by a denser region. We searched a $\sim 0.25$ deg$^2$ area to the south of Cyg X--1 in \o3, and a $\sim 1$ deg$^2$ H$\alpha$ mosaic from the INT Photometric H$\alpha$ Survey of the Northern Galactic Plane (IPHAS) \cite{drewet05}. Two H$\alpha$ candidate hot spots analogous with those associated with AGN lobes could be confirmed with follow-up observations in radio, \s2 (6725\AA; which is an emission line diagnostic of shocked gas as opposed to photoionized gas \cite{oste89}) or deeper observations in \o3.

\section{The Galactic survey for microquasar jet-powered nebulae}

\begin{figure}
\begin{center}
\includegraphics[width=12cm,angle=0]{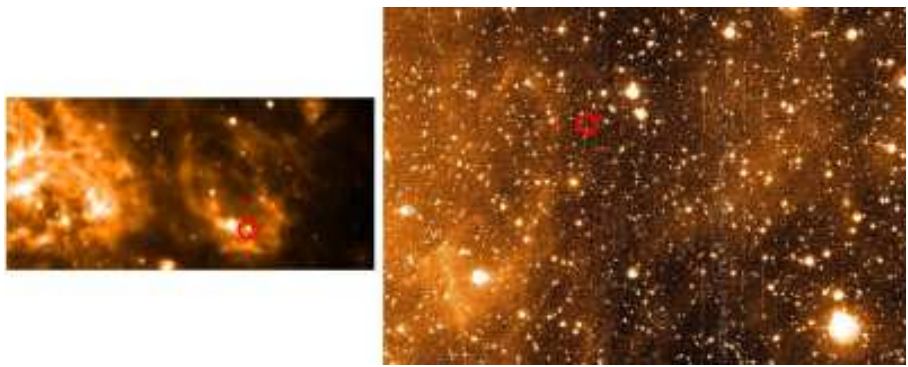}
\caption{H$\alpha$ images of the regions local to LMC X--1 (left) and GRS 1009--45 (right). North is up and east is to the left. The X-ray binaries are marked in red. An X-ray ionized nebula (XIN) surrounds LMC X--1 \cite{pakuan86} and a shell-like structure lies to the south-east of GRS 1009--45. Follow-up observations are required to identify the nature of this latter nebula. The jets of LMC X--1 may be interacting with the XIN which may be visible via enhanced [S II] emission.}
\end{center}
\end{figure}

Are there jet-powered nebulae associated with other microquasars? If so, their properties will test the ubiquity of the jet power from microquasars and will further constrain the matter and energy microquasars input into the ISM. Low-mass X-ray binaries may have less powerful jets than Cyg X--1 when time-averaged, due to the long phases in quiescence, but they are longer lived and so the [jet power $\times$ lifetime] product may be comparable. Any nebulae discovered associated with the jet of a low-mass X-ray binary will provide constraints on these speculations.

We were awarded time on two telescopes for this project: the Isaac Newton Telescope (on La Palma; northern hemisphere) and the ESO/MPI 2.2m telescope (in Chile; southern hemisphere). 0.3 deg$^2$ fields around 30 X-ray binaries were observed in H$\alpha$. In most fields no shell-like nebulae were discovered, however a few candidates were found.

In Fig. 3 we show H$\alpha$ images of the regions around LMC X--1 and GRS 1009--45. A known X-ray-ionized nebula surrounds LMC X--1 which is photoionized by the UV and X-ray photons from the X-ray binary \cite{pakuan86}. Follow-up observations in (for example) \s2 may reveal regions of the nebula where the jets are interacting with the photoionized gas (as \s2 is much stronger in shocked gas than photoionized gas). For GRS 1009--45, an oval-shaped nebula to the east of the X-ray binary may be shock-excited from the jets, and is consistent with the morphology of some AGN radio lobes. Follow-up narrowband observations are required to test this.

\begin{figure}
\begin{center}
\includegraphics[width=15cm,angle=0]{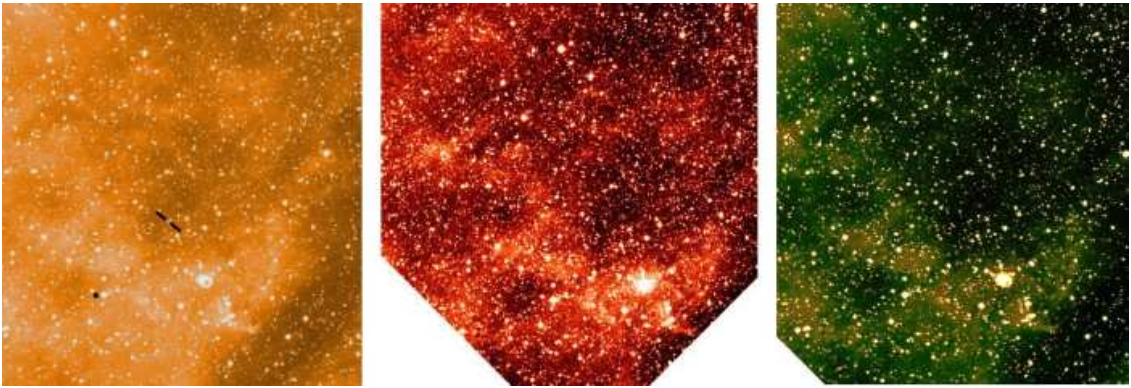}
\caption{GRO J1655--40 (marked black in the left panel; the direction is that of the resolved radio jets \cite{hjelru95}) and its candidate jet-powered nebula to the lower right in each panel. North is up and east is to the left. From left to right: H$\alpha$ image taken with the ESO/MPI 2.2~m; [S II] image taken with the Danish 1.5~m; [S II] / H$\alpha$ ratio image (H$\alpha$ is green, [S II] is red).}
\end{center}
\end{figure}

\subsection{GRO J1655--40: The best candidate}

An apparent H$\alpha$ shell to the south west of GRO J1655--40 is aligned with the resolved radio jet \cite{hjelru95}. We performed follow-up imaging in the \s2 filter with the Danish 1.5m telescope. The shell is clearly visible in \s2 line emission implying the gas is shocked-excited as opposed to photoionized (Fig. 4). There is also extended radio emission in the region of this nebula \cite{martet96}. Flux calibration of the optical images and overlaying of archival radio data \emph{may confirm this structure to be the second microquasar jet-powered nebula}.

To summarise this section, we do not find many Galactic jet-powered nebulae, probably because the conditions required are rare: a high local mass density is required for a bow shock to form \cite{hein02}, however low Galactic extinction towards the source is also needed to see these faint structures, and a low space velocity of the system may be needed so that the power is not
dissipated over too large a volume.

\section{Extragalactic candidates}

\begin{figure}
\begin{center}
\includegraphics[width=10cm,angle=0]{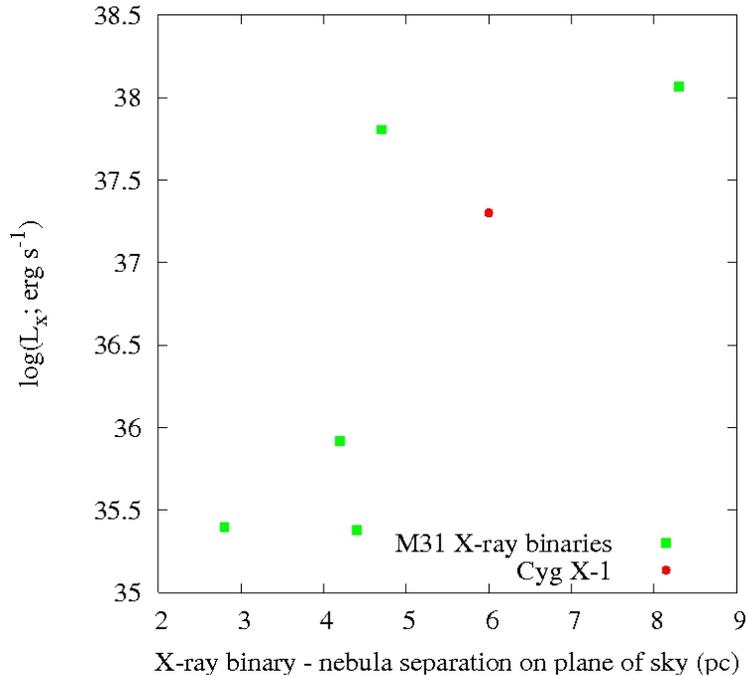}
\caption{The separation on the plane of the sky between the emission line sources in M31 and their nearby X-ray binaries, versus the time-averaged X-ray luminosity of the X-ray binary. Cyg X--1 and its jet-powered nebula is also plotted. The inclination angle of the jet axis to the line of sight plays a role in varying the projected separation on the sky plane, which will invalidate any apparent relation between the separation on the plane of the sky with $L_{\rm X}$.}
\end{center}
\end{figure}

The conditions suggested at the end of the last section could be met for extragalactic X-ray binaries. Williams et al. (2004) found six X-ray binaries in M31 peculiarly close to strong \o3 emission line `planetary nebulae' \cite{willet04}. The chance of this proximity is $\sim 1$ percent so the objects are very likely to be associated with the binary systems. Five of the six have strong \s2 emission ($F_{\rm \s2}$/$F_{\rm H\alpha} \geq 1$, and so originate in shocked gas; they are therefore much more likely to be supernova remnants (SNRs) than planetary nebulae. The X-ray binaries cannot have formed from the progenitors of the SNRs because the kick velocities required for the observed separations ($>3000$ km s$^{-1}$) would disrupt the binaries.

\begin{figure}
\begin{center}
\includegraphics[width=12cm,angle=0]{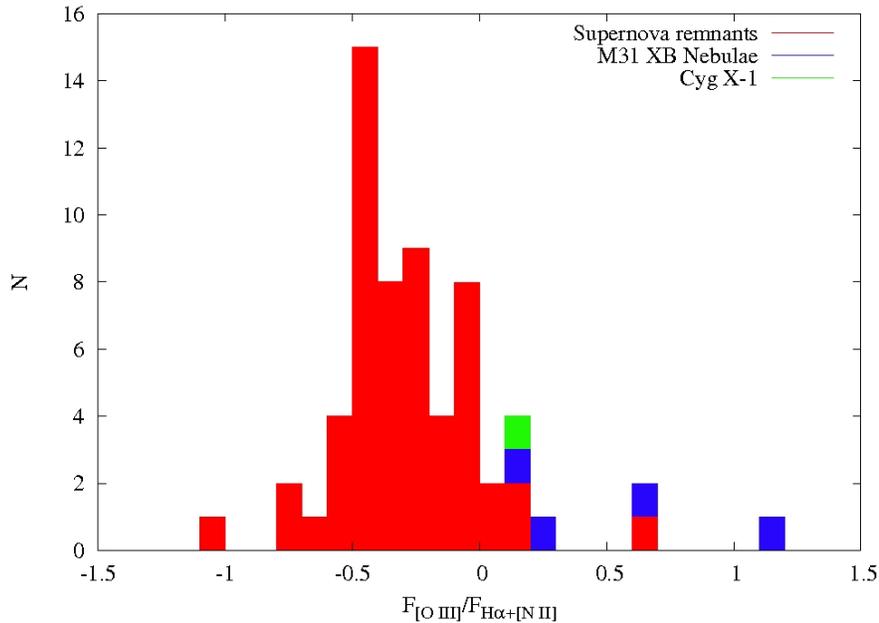}
\caption{A histogram representing the distribution of [O III] / H$\alpha$ ratios in planetary nebulae and in the candidate X-ray binary jet-powered nebulae in M31 (and the Cyg X--1 jet-powered nebula).}
\end{center}
\end{figure}

We suggest that these emission line sources could be nebulae powered by the jets of the X-ray binaries. The separation on the plane of the sky between the candidate nebulae and associated X-ray binaries, and the time-averaged X-ray luminosity, are of the same order as the Cyg X--1 nebula and Cyg X--1 (Fig. 5). The \o3 / H$\alpha$ ratio is also higher in general for these nebulae than SNRs taken from the literature \cite{oste89,dopiet80,blaiet82,feseet85,matoet97,mavret02} (Fig. 6) - could this be a diagnostic for separating jet-powered nebulae from SNRs? The higher \o3 / H$\alpha$ ratio could be explained by faster shock velocities due to the more powerful directional energy input for microquasar jets than for SNRs. Photoionized nebulae like H~\small II \normalsize regions and planetary nebulae can be separated from SNRs and jet-powered nebulae by the \s2 / H$\alpha$ ratio \cite{feseet85}.

In addition, it is interesting to note that many ultraluminous X-ray sources (ULXs) are surrounded by emission nebulae or \emph{supershells} with diameters $\geq$100 pc and some are bipolar \cite{pakumi02}. Perhaps these could be powered by winds or jets from the ULXs \cite{pakumi02} or photoionized by the X-ray source, like the nebula surrounding LMC X--1 \cite{pakuan86}.

\section{Summary}

Analysing jet-ISM interactions may provide more accurate measurements of the time-averaged jet power than available from the core radio luminosity. The jet from Cyg X--1 is producing a bow shock that is ploughing through the ISM at $>100$ km s$^{-1}$. We have identified for the first time the high temperature bow shock front powered by an X-ray binary jet. The total power inferred from both jets is $\sim$0.3--0.8 times the time-averaged Bolometric X-ray luminosity of Cyg X--1 \cite{russet06b}. GRO J1655--40, GRS 1009--45 and five X-ray binaries in M31 have candidate jet-powered nebulae. Follow-up observations are required to confirm these candidates. It may be possible to distinguish jet-powered nebulae from other nebulae (SNRs, planetary nebulae, H~\small II \normalsize regions) via their optical emission line ratios.

\vspace{5mm}
\emph{Acknowledgements}
This paper makes use of data from the Isaac Newton Telescope, operated on the island of La Palma by the Isaac Newton Group in the Spanish Observatorio del Roque de los Muchachos of the Instituto de Astrofisica de Canarias. Also based on observations made with the Danish 1.54-m and ESO 2.2-m (under ESO programme ID 076.D-0017) Telescopes at the La Silla Observatory. We are grateful to Klaas Wiersema for obtaining the \s2 data of GRO J1655--40 with the Danish 1.54-m. EG is supported by NASA through Chandra Postdoctoral Fellowship Award PF5-60037, issued by the Chandra X-Ray Observatory Center, which is operated by the Smithsonian Astrophysical Observatory for and on behalf of NASA under contract NAS8-39073.

\end{document}